\documentclass[conference]{IEEEtran}
\IEEEoverridecommandlockouts
\usepackage{cite}
\usepackage{amsmath,amssymb,amsfonts}
\usepackage{algorithmic}
\usepackage{graphicx}
\usepackage{textcomp}
\def\BibTeX{{\rm B\kern-.05em{\sc i\kern-.025em b}\kern-.08em
    T\kern-.1667em\lower.7ex\hbox{E}\kern-.125emX}}

\usepackage{comment}

\usepackage{booktabs, multirow}
\usepackage[table,xcdraw]{xcolor}
\usepackage{svg}

\begin{document}

\title{Emotion-Aware Prefix: Towards Explicit Emotion Control in Voice Conversion Models
}


\author{
\IEEEauthorblockN{Haoyuan Yang, Mu Yang, Jiamin Xie, Szu-Jui Chen, John H.L. Hansen}
\IEEEauthorblockA{Center for Robust Speech Systems (CRSS), The University of Texas at Dallas, USA \\
\{haoyuan.yang, mu.yang, jiamin.xie, szu-jui.chen, john.hansen\}@utdallas.edu}
}


\maketitle

\begin{abstract}
    Recent advances in zero-shot voice conversion have shown potential for emotion control, yet their explicit emotion controllability remains limited and inconsistent. We propose Emotion-Aware Prefix for explicit emotion control in a two-stage voice conversion backbone. The proposed method substantially improves emotion conversion performance, nearly doubling the baseline Emotion Conversion Accuracy (ECA) from 47.5\% to 91.2\%, while preserving linguistic content, speech quality, and speaker identity. Ablation studies show that the emotion and style pathways affect different aspects of the output; stage-wise analyses further indicate that sequence modulation is the primary driver of emotion control, while acoustic realization complements it for reliable conversion.Finally, comparative analysis suggests that acoustic decoupling may help maintain speaker identity under strong emotion control.
    \footnote{Demos available at: https://anonysample.github.io/vevoemoplus/} \footnote{This work was supported by a grant from IARPA-ARTS, and partially by the Univ. of Texas – Dallas from the Distinguished University Chair in Telecommunications Engineering held by J.H.L. Hansen.}
\end{abstract}



\begin{figure*}[t] 
    \centering
    \includegraphics[width=0.95\linewidth]{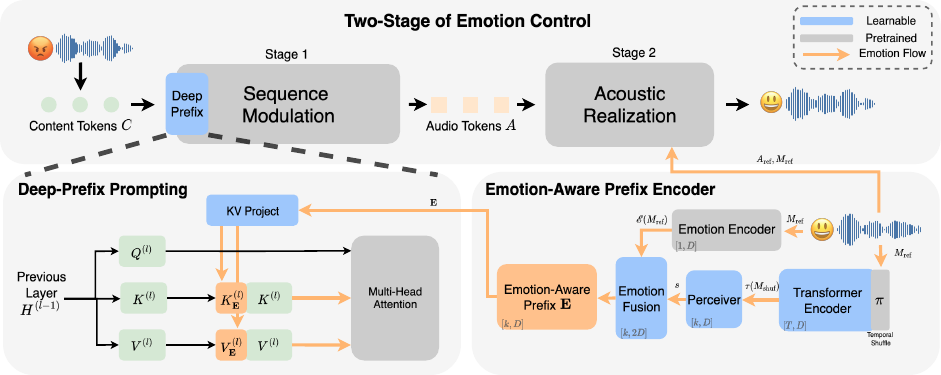}
    \caption{Framework Overview: Sequence Modulation Stage performs sequence-level style modeling via Deep-Prefix Prompting of Emotion-Aware Prefix, while Acoustic Realization Stage models the final speech production conditioned on reference speech.}
    \label{fig:architecture}
\end{figure*}
\section{Introduction}
Emotion control is essential for the naturalness and liveliness of speech generation, as it conveys a speaker's feelings, motivations, and personality~\cite{Scherer2003VocalCO}. 
Effective emotion control is a cornerstone for creating truly immersive Human-Computer Interfaces, enabling applications ranging from expressive dubbing and human-computer interaction to speaker anonymization~\cite{DBLP:conf/icassp/ZhaoL025b/dubbing, DBLP:conf/interspeech/YaoLCX25/easy}.

Recently, zero-shot voice conversion has emerged as a powerful paradigm that inherently possesses the potential for emotion conversion. By exploiting style information from a reference prompt, these models convert a linguistic content sequence into a style-rich audio token sequence, followed by acoustic modeling to reconstruct speech with the target identity and style~\cite{ DBLP:journals/corr/abs-2502-04519/genvc}. Notably, VEVO proposes a two-stage design that decomposes zero-shot voice conversion into sequence modulation and acoustic realization, which separates high-level temporal and prosodic structures from low-level spectral synthesis~\cite{DBLP:conf/iclr/ZhangZPTMLHLWCH25/vevo}. This resonates with the multi-scale organization of speech production, where dynamic modulation of prosodic patterns operating within relatively stable, speaker-specific acoustic constraints~ realizes emotional expressions~\cite{scherer1986vocal}.

However, our study finds that existing zero-shot voice conversion models often exhibit suboptimal or inconsistent emotion controllability. While they can mimic the overall speaking style, they lack the control required to shift a source utterance into a specific, high-intensity target emotion. This limitation stems from lacking explicit control for emotion during the dynamic modulation stage, leaving the models overly reliant on implicit cues (e.g., global energy or average pitch) given by the acoustic prompt.

In this paper, we propose Emotion-Aware Prefix for explicit emotion control in voice conversion models. Tailored for the emotional voice conversion (EVC) task, our approach enables explicit emotion control while preserving linguistic content, speaker identity and overall quality. Our proposed method is based on VEVO \cite{DBLP:conf/iclr/ZhangZPTMLHLWCH25/vevo}, a two-stage zero-shot voice conversion framework. As illustrated in Figure~\ref{fig:architecture}, we leverage the emotion cues within prompt speech throughout the two-stage architecture. By integrating content-invariant style features with an emotion vector, we develop an Emotion-Aware Prefix. This prefix directs the sequence modulation stage through a Deep-Prefix Prompting mechanism, ensuring explicit control over the generated audio tokens. Simultaneously, the same reference is utilized during the acoustic realization stage to provide stable, low-level spectral information. Our main contributions are: 



\begin{enumerate}
    \item \textbf{Enhancing Emotion Controllability}. By introducing Emotion-Aware Prefix, we improve the emotion controllability of VEVO, increasing target Emotion Conversion Accuracy (ECA) from 47.5\% to 91.2\% while maintaining speaker identity and speech quality.

    \item \textbf{Understanding Hierarchical Sensitivity in Emotion Control}. Through stage-wise emotion-prompt isolation, we show that sequence-level modulation serves as the primary driver of high-level prosodic intent, while joint control with the acoustic stage yields a large non-additive improvement in conversion accuracy.

    \item \textbf{Investigating the Role of Acoustic Decoupling}. We apply the proposed method to a voice conversion framework without acoustic decoupling. The comparative analysis shows that the proposed module can improve emotion control in another backbone, while suggesting that acoustic decoupling may help preserve speaker identity under strong emotion control.
\end{enumerate}

\section{Related Work}

Early work on expressive speech modification explored transforming neutral speech into stressed speaking styles through explicit acoustic perturbation and parametric synthesis~\cite{bou1996stressed,bou1996stress}, while modern EVC focuses more specifically on manipulating emotional states while preserving speaker identity. Early approaches primarily relied on explicit representation disentanglement, employing dedicated emotion encoders~\cite{DBLP:conf/interspeech/HeCRS21/stargan-evc-improved} or mutual information objectives~\cite{DBLP:conf/interspeech/DuSZ022/disentangle-evc} to separate emotional style from speaker identity and linguistic content. To mitigate reliance on high-quality parallel data~\cite{DBLP:journals/speech/ZhouSLL22/esd} and extensive human annotation~\cite{DBLP:journals/corr/abs-2509-09791/msp}, subsequent research shifted toward non-parallel~\cite{DBLP:conf/icassp/RizosBES20/stargan-evc,DBLP:conf/interspeech/ZhuZ0023/semi-evc} and in-the-wild settings~\cite{DBLP:conf/interspeech/ChouLSTL25/zsdevc}. More recent works have further moved beyond rigid categorical emotion transfer toward nuanced expressiveness, including controllable speech emotion generation~\cite{DBLP:journals/taffco/ChoOKL25/emosphere-plus} and fine-grained intensity regulation~\cite{DBLP:conf/aaai/GudmalwarBSWS25/emoreg,emovox}. Another closely related line specifies the target emotion through reference speech or natural-language prompts, including zero-shot emotion style transfer~\cite{zest}, reference-based emotion conversion~\cite{emovox}, and prompt-based EVC systems~\cite{clapfm_evc,promptevc}. These studies highlight the importance of explicit target-emotion conditioning. Kreuk et al.~\cite{DBLP:conf/emnlp/KreukPCKNRHMDA22/s2s-evc} demonstrated that emotion can be manipulated through discrete acoustic token representations, suggesting that emotional attributes can be implicitly encoded and controlled within autoregressive sequence models. Building on this direction, recent voice conversion backbones based on discrete tokens~\cite{DBLP:journals/corr/abs-2502-04519/genvc,DBLP:conf/iclr/ZhangZPTMLHLWCH25/vevo} provide a natural testbed for studying how explicit emotion conditioning should be injected into two-stage VC models.

\begin{table*}[t]
\centering
\caption{Objective evaluation results. The proposed method improves emotion conversion accuracy while maintaining speaker, quality, and intelligibility metrics. Emotion is evaluated using a zero-shot emotion2vec recognizer (e2v) and an in-domain wav2vec2 SER recognizer (w2v SER). Rows below the divider report the proposed method, analysis probes and ablations.}
\label{tab:main_results}
\setlength{\tabcolsep}{6pt}
\begin{tabular}{l|cc|cccc|c|cc}
\toprule
\multirow{3}{*}{\textbf{Model}} & \multicolumn{2}{c|}{\textbf{Speaker}} & \multicolumn{4}{c|}{\textbf{Quality \& Intelligibility}} & \multicolumn{3}{c}{\textbf{Emotion}} \\
\cmidrule(lr){2-3}\cmidrule(lr){4-7}\cmidrule(lr){8-10}
& \multirow{2}{*}{Spk-Cent SIM $\uparrow$} & \multirow{2}{*}{EER $\downarrow$} & \multirow{2}{*}{UT-MOSv2 $\uparrow$} & \multicolumn{2}{c}{DMOS} & \multirow{2}{*}{WER $\downarrow$} & \textbf{e2v} & \multicolumn{2}{c}{\textbf{w2v SER}} \\
\cmidrule(lr){5-6}\cmidrule(lr){8-8}\cmidrule(lr){9-10}
& & & & SIG & OVRL & & ECA $\uparrow$ & Emo SIM $\uparrow$ & ECA $\uparrow$ \\
\midrule
StarGANv2-VC-EVC & \textbf{0.565} & \textbf{3.24\%} & 2.318 & 3.330 & 3.037 & 5.87\% & 36.0\% & 0.405 & 36.8\% \\
GenVC & 0.238 & 20.87\% & 2.157 & 3.328 & 3.021 & 8.06\% & 32.5\% & 0.321 & 32.2\% \\
StepAudioEditX (iter 2) & 0.425 & 6.50\% & \textbf{3.395} & 3.308 & 3.005 & \textbf{4.43\%} & 20.9\% & 0.241 & 21.9\% \\
VEVO & 0.476 & 5.40\% & 3.060 & 3.409 & 3.093 & 10.08\% & 42.4\% & 0.536 & 47.5\% \\
\midrule
\textbf{Proposed} & 0.500 & 4.50\% & 2.960 & \textbf{3.431} & \textbf{3.152} & 6.28\% & \textbf{85.5\%} & \textbf{0.866} & \textbf{91.2\%} \\
\quad AR-LoRA only & 0.508 & 5.04\% & 2.989 & 3.312 & 3.002 & 8.13\% & 70.0\% & 0.802 & 73.2\% \\
\quad w/o Target Emo Emb. & 0.491 & 3.29\% & 2.846 & 3.283 & 2.974 & 7.81\% & 44.4\% & 0.525 & 43.4\% \\
\quad w/o Target Style Emb. & 0.464 & 4.33\% & 2.838 & 3.278 & 2.965 & 7.77\% & 77.3\% & 0.846 & 86.1\% \\
\quad w/o Deep-Prefix & 0.503 & 4.42\% & 2.946 & 3.416 & 3.133 & 5.74\% & 83.5\% & 0.862 & 90.6\% \\
\bottomrule
\end{tabular}
\end{table*}

\section{Method}
To enhance the expressiveness of emotion with minimal architectural modification, we extend VEVO \cite{DBLP:conf/iclr/ZhangZPTMLHLWCH25/vevo}, a state-of-the-art voice conversion framework with \textbf{Emotion-Aware Prefix} and Deep-Prefix prompting.

\subsection{Framework Overview}
As shown in Figure~\ref{fig:architecture}, our proposed framework follows the two-stage speech modeling in VEVO:

\noindent\textbf{Stage 1: Sequence Modulation}. 
Stage 1 employs an Autoregressive (AR) Transformer to predict discrete, style-rich audio tokens. The prompt to the AR transformer combines the given content tokens and the proposed emotion-aware prefix $\mathbf{E}$, which will be discussed in section~\ref{EAP}. This process is formulated as:
\begin{align}
P(A | C, \mathbf{E}) = \prod_{t=1}^{T} P(a_t | a_{<t}, C, \mathbf{E})
\end{align}
where $A=[a_1, a_2, ..., a_T]$ are the audio tokens to be predicted and $C =[c_1, c_2, ..., c_n]$ denotes content tokens extracted from input speech. 


\noindent\textbf{Stage 2: Acoustic Realization}. 
Stage 2 employs a Flow-Matching (FM) Transformer to reconstruct the mel-spectrogram $\widehat{\mathrm{M}}$ from the predicted audio tokens $A$, conditioned on reference audio tokens $A_{\text{ref}}$ and their corresponding mel-spectrogram $\mathrm{M}_{\text{ref}}$:
\begin{align}
\widehat{\mathrm{M}} = \mathcal{F}_{\theta}\!\left(\epsilon; A, A_{\text{ref}}, \mathrm{M}_{\text{ref}}\right), \quad \epsilon \sim \mathcal{N}(0, I),
\end{align}
where $\mathcal{F}_{\theta}$ is the solution operator of the conditional flow-matching ODE parameterized by the FM Transformer, and the final waveform is produced by a neural vocoder. We keep this stage frozen: the target emotion enters Stage~2 only through the reference $(A_{\text{ref}}, \mathrm{M}_{\text{ref}})$, drawn from the target speaker in the target emotion. This preserves the acoustic decoupling of VEVO and, as analyzed in Section~\ref{sec:results}, appears important for retaining speaker identity under strong emotion control.



\subsection{Emotion-Aware Prefix Encoder}
\label{EAP}
The Emotion-Aware prefix encoder provides utterance-level, content-invariant emotion style embedding.
The encoder consists of a Temporal-Shuffle Transformer, a Perceiver Layer, and an Emotion Fusion Layer. Given a reference mel-spectrogram $\mathrm{M}_\text{ref}\in \mathbb{R}^{T \times D}$, these modules encode $\mathrm{M}_\text{ref}$ as follows:

\noindent\textbf{Temporal-Shuffle Transformer Encoder}. 
To reduce content leakage while preserving global style characteristics, we apply a random permutation $\pi(\cdot)$ to the temporal indices of the reference mel-spectrogram, yielding $\mathrm{M}_{\text{shuf}} = \pi(\mathrm{M}_{\text{ref}})$. By disrupting temporal order, this operation suppresses time-aligned phonetic and linguistic structure and discourages content leakage, while retaining order-invariant global characteristics such as timbre and overall spectral statistics. Time-varying emotional dynamics are then generated by the sequence modulation stage under explicit emotion conditioning. The shuffled frames are then processed by a lightweight Transformer $\mathcal{T}$ to extract high-level style representations.

\noindent\textbf{Perceiver Layer}. Inspired by the architecture of GenVC\cite{DBLP:journals/corr/abs-2502-04519/genvc}, we employ a Perceiver layer to compress the variable-length latent features $\mathcal{T}(\mathrm{M}_\text{shuf})$ into a fixed-length style embedding $s \in \mathbb{R}^{k \times D}$. This is achieved by fixing $k$ learnable latent tokens $L$ that cross-attend to $\mathcal{T}(\mathrm{M}_\text{shuf})$. The fixed-length design works as a bottleneck to ensure consistent dimensionality for prefix injection into the AR Transformer, preventing style representation from scaling with utterance duration.

\noindent\textbf{Emotion Fusion Layer}. To incorporate explicit emotion information, a pre-trained emotion encoder $\mathcal{E}$ extracts an embedding from the reference mel-spectrogram $\mathrm{M}_\text{ref}$. This is then broadcast and fused with the style embeddings $s$ and projected to the hidden dimension of the language model: 
\begin{align}
    \mathbf{E} = \text{Linear}(\text{Concat}[s ; \mathcal{E}(\mathrm{M}_\text{ref})])
\end{align}
where $\text{Concat[;]}$ denotes concatenation in the feature dimension. The resulting vector $\mathbf{E}$ serves as the \textbf{Emotion-Aware Prefix} that enhances the subsequent audio token generation.

\subsection{Deep-Prefix Prompting}
To ensure consistent emotion control in the sequence modulation stage throughout the duration of the generated tokens, we employ a Deep-Prefix Prompting mechanism similar to P-Tuning-v2~\cite{Liu2021PTuningVP/ptuning}. Rather than simply prepending the Emotion-Aware Prefix $\mathbf{E}$ to the input sequence, we inject it as the layerwise KV-cache of the language model.
At each layer, we utilize independent key and value projections to map the emotion-aware prefix $\mathbf{E}$ to the latent space of that layer:
\begin{align}
    K_{\mathbf{E}}^{(l)} = \mathbf{E} \cdot W_K^{(l)}, \quad V_{\mathbf{E}}^{(l)} = \mathbf{E} \cdot W_V^{(l)}
\end{align}
where $W_K^{(l)}$ and $W_V^{(l)}$ denote the key and value projection matrix for layer $l$ of the language model, respectively.
The resulting prefix vectors $K_{\mathbf{E}}^{(l)}$ and $V_{\mathbf{E}}^{(l)}$ are prepended to the standard key and value matrices derived from the hidden states of the previous layer as $[K^{(l)}_{\mathbf{E}}; K^{(l)}]$ and $[V^{(l)}_{\mathbf{E}}; V^{(l)}]$, followed by a standard attention computation.

\noindent\textbf{Training.} Only the Emotion-Aware Prefix encoder and a lightweight adapter on the AR Transformer are trainable; the backbone and the entire acoustic stage remain frozen. We optimize the standard next-token prediction objective of Stage~1 and introduce no explicit emotion classification or alignment loss---emotion control is learned through prefix conditioning under the next-token prediction objective. This design avoids any supervision that directly targets the emotion metrics, and keeps Stage~2 untouched as a fixed identity anchor.

\begin{figure*}[t] 
    \centering
    \includegraphics[width=0.90\linewidth]{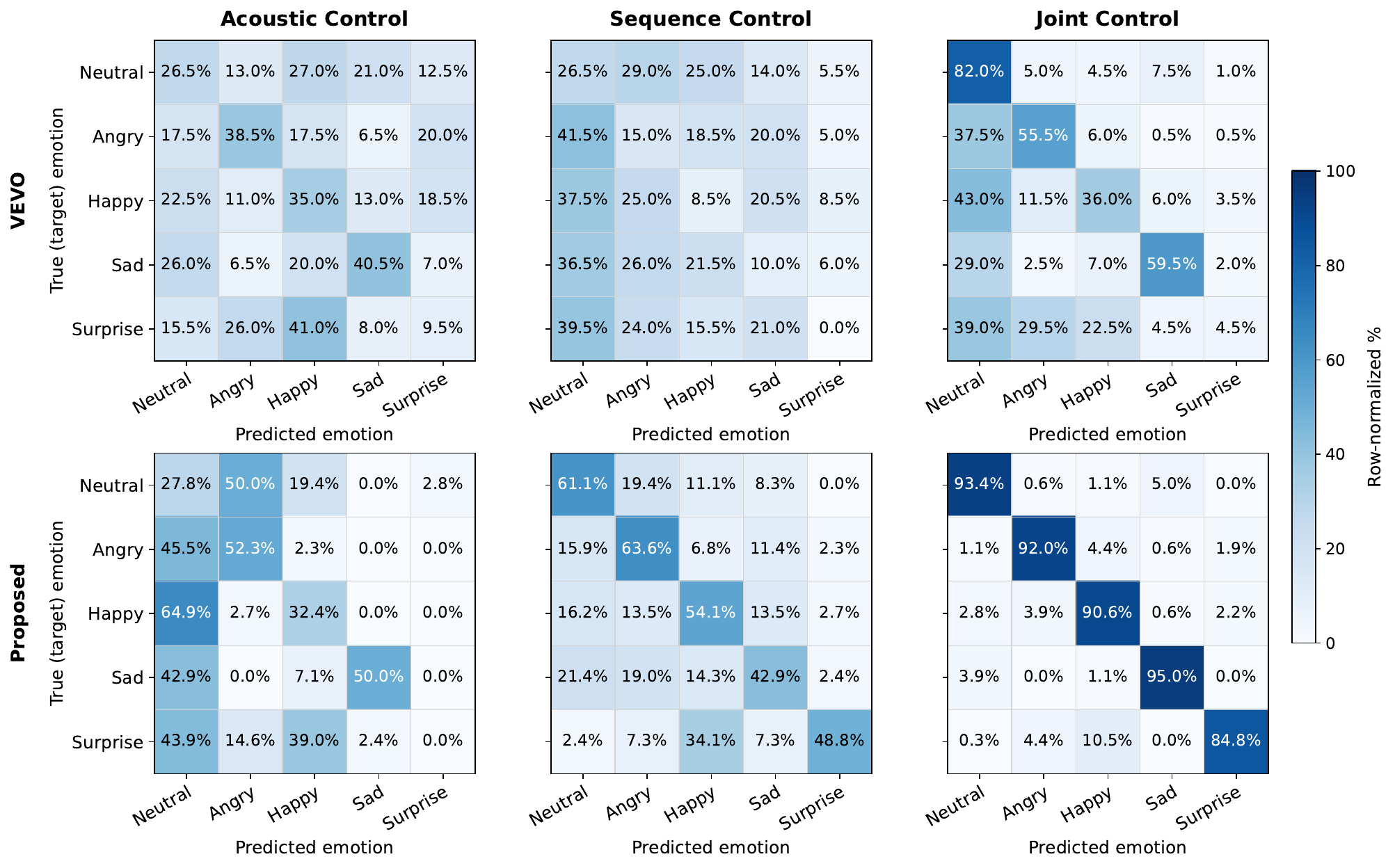}
    \caption{Per-emotion confusion matrices for stage-wise emotion control. Rows denote target emotions and columns denote w2v SER predictions; values are row-normalized percentages. Acoustic, Sequence, and Joint control indicate target-emotion prompting of Stage 2 only, Stage 1 only, and both stages, respectively.}
    \label{fig:confusion}
\end{figure*}

\section{Experiments}
\subsection{Dataset}
We train the proposed model on Emotional Speech Database (ESD)\cite{DBLP:journals/speech/ZhouSLL22/esd}. We select all 10 English speakers (5 male, 5 female) across all five emotions (Neutral, Happy, Sad, Angry and Surprise). For each speaker and emotion, the dataset provides 350 utterances in parallel. We utilize the first 300 utterances for training, with the remaining 50 split into 20 for reference prompts and 30 for testing.

\subsection{Implementation Details}
We build on VEVO with its original tokenizers, AR/FM Transformers, and vocoder. The Emotion-Aware Prefix encoder uses a 6-layer Temporal-Shuffle Transformer and a Perceiver layer with $k=32$ latents, and a pretrained Emotion2Vec+ Large~\cite{DBLP:conf/acl/MaZYLGZ024/emo2vec} model as the emotion encoder $\mathcal{E}$. The adapter on the AR Transformer is LoRA~\cite{DBLP:conf/iclr/HuSWALWWC22/LoRA} with rank $r=32$. We train for 46k steps with AdamW at a learning rate of $2 \times 10^{-5}$.

\subsection{Baselines}
We compare our results against four prominent baseline systems: (1) \textbf{VEVO}, a state-of-the-art zero-shot voice conversion system known for high-fidelity style mimicry, also the backbone of the proposed method; (2) \textbf{GenVC}, a zero-shot voice conversion model that utilizes a Perceiver encoder to disentangle speaker style from content without supervised labels; (3) \textbf{StarGANv2-VC-EVC}: our customized implementation of an EVC variant of StarGANv2-VC \cite{DBLP:conf/interspeech/LiZM21/starganv2-vc-evc}, following the principles of StarGAN-EVC for many-to-many emotion transfer; (4) \textbf{StepAudioEditX}~\cite{yan2025stepaudioeditxtechnicalreport/stepaudio}: a 3B-parameter Large Audio Language Model (LALM) serving as a benchmark for foundation-model-based expressive audio editing, for which we report the results of the second editing iteration, denoted as iter 2.

\subsection{Objective Evaluation}
We evaluate the performance across three main dimensions:

\noindent \textbf{Speaker}. Speaker embedding is extracted using ECAPA-TDNN~\cite{DBLP:conf/interspeech/DesplanquesTD20}. Following the observations that intra-speaker similarity may vary across emotional conditions~\cite{DBLP:conf/odyssey/DuL0KS24/devc}, we compute a speaker-level cross-emotion embedding centroid for each speaker using the training data and use them as identity anchors. We report Speaker Centroid Similarity (Spk-Cent SIM), defined as cosine similarity between converted utterances and target speaker centroids, along with Equal Error Rates (EER) following a standard speaker verification task.

\noindent \textbf{Quality and Intelligibility}. We use UTMOSv2~\cite{baba2024utmosv2/utmos} to assess naturalness, and DNSMOS P.835~\cite{DBLP:conf/icassp/ReddyGC22/dnsmos} for overall perceptual quality. 
For intelligibility, we use Whisper-v3-Large~\cite{radford2022whisper/whisper} to transcribe the generated speech and compute Word Error Rate (WER) against the ground truth text.

\noindent \textbf{Emotion}. We report Emotion Conversion Accuracy (ECA), the top-1 agreement between the predicted emotion of a converted utterance and its target label, and Emotion Similarity (Emo SIM), the cosine similarity between converted and target emotion embeddings. We evaluate emotion with two models:
\textbf{(i)~emotion2vec}~\cite{DBLP:conf/acl/MaZYLGZ024/emo2vec}: a SER model that is \emph{zero-shot} for our task and is not fine-tuned on ESD.  Considering that it is also used as the feature extractor $\mathcal{E}$ during training, we only report the recognition accuracy without computing the embedding similarity to avoid circularity in embedding-based evaluation. \textbf{(ii)~w2v SER}: a wav2vec2-large-xlsr~\cite{conneau21_xlsr} based SER model fine-tuned on SAVEE~\cite{savee}, RAVDESS~\cite{livingstone2018ryerson}, TESS~\cite{tess}, and ESD training split, reaching 92.6\% accuracy on the ESD test set. The model retains its original seven-class head; predictions falling outside the five ESD emotions (e.g.\ disgust or fear) are counted as errors rather than being remapped, yielding a stricter evaluation. This model is \emph{in-domain} but is not used to train our voice conversion model. We therefore treat w2v SER as the primary objective emotion metric and report emotion2vec as a zero-shot cross-check. In Table~\ref{tab:main_results}, ECA is reported for both recognizers, while Emo SIM is computed only from w2v SER embeddings. Unless otherwise noted, the analysis refers to the w2v SER results.

    


\subsection{Subjective Evaluation}

\begin{figure}[t] 
    \centering
    \includegraphics[width=1.0\linewidth]{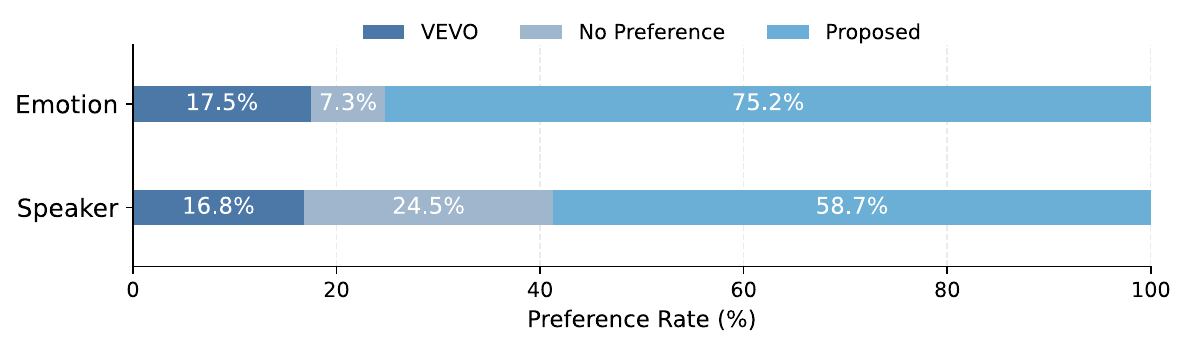}
    \caption{Subjective preference between VEVO and the proposed method. The proposed method is preferred in 75.2\% of emotion-similarity trials (Wilson 95\% CI: [70.8, 79.2]\%) and 58.7\% of speaker-similarity comparisons (Wilson 95\% CI: [53.9, 63.5]\%).}
    \label{fig:listener} 
\end{figure}

Ten listeners (five male, five female) evaluated speech quality, speaker similarity, and emotion similarity on samples from the
VEVO baseline and our proposed method. Each listener completed 40 trials per dimension, for a total of 120 trials per listener. This yielded 400 listener-trials for each evaluation dimension and 1,200 listener-trials overall. To prevent listeners from inferring system identity, the baseline and proposed samples were randomly assigned to the A/B positions in every trial.

\noindent\textbf{Quality}. Participants evaluated the overall quality of converted samples using a rating scale from 1 (poor) to 5 (excellent). We report the Mean Opinion Score (MOS).

\noindent\textbf{Emotion Preference}. Subjective emotion similarity was evaluated using an ABX preference test. In each trial, listeners were presented with a reference utterance and two converted samples (baseline and proposed). They selected the sample whose emotion was closer to the reference, or selected no preference when no clear difference was perceived. We report the preference rate.

\noindent\textbf{Speaker Preference}. Subjective speaker similarity was assessed using a 5-point rating scale (1: very dissimilar, 5: highly similar) with respect to the reference utterance. For each trial, preference statistics were derived by comparing the speaker similarity scores assigned to the baseline and proposed samples. The system with the higher score was counted as preferred, while equal scores were counted as no preference. We report the score-derived preference rate.

For preference statistics, no-preference responses were included in the denominator. The 95\% confidence intervals for preference rates were computed using the Wilson score interval over all 400 listener-trials in each evaluation dimension.



\begin{table}[t]
\centering
\caption{Stage-wise emotion control evaluated with the w2v SER. Acoustic Only, Sequence Only, and Joint denote target-emotion prompting of Stage~2 only, Stage~1 only, and both stages, respectively.}
\label{tab:hierarchical_sensitivity}
\setlength{\tabcolsep}{8pt}
\begin{tabular}{ll|c|cc}
\toprule
\textbf{Control} & \textbf{Model} & \textbf{EER} $\downarrow$ & \textbf{Emo SIM} $\uparrow$ & \textbf{ECA} $\uparrow$ \\
\midrule
\multirow{2}{*}{Acoustic Only} & VEVO & 4.50\% & 0.402 & 30.0\% \\
                               & Proposed & 3.53\% & 0.438 & 33.0\% \\
\midrule
\multirow{2}{*}{Sequence Only} & VEVO & 3.98\% & 0.132 & 12.0\% \\
                               & Proposed & \textbf{3.00\%} & 0.537 & 54.0\% \\
\midrule
\multirow{2}{*}{\textbf{Joint}} & VEVO & 5.40\% & 0.536 & 47.5\% \\
                               & Proposed & 4.50\% & \textbf{0.866} & \textbf{91.2\%} \\
\bottomrule
\end{tabular}
\end{table}

\section{Results and Analyses}
\label{sec:results}
\subsection{Effects of Emotion-Aware Prefix}
\noindent\textbf{Objective Evaluation}.

Table~\ref{tab:main_results} reports the proposed method against the baselines and ablations. The improvement is consistent across both emotion recognizers: ECA increases from 42.4\% to 85.5\% under the zero-shot emotion2vec recognizer and from 47.5\% to 91.2\% under the in-domain w2v SER. The AR-LoRA-only variant, which removes the Stage-1 EAP while keeping the same Stage-2 acoustic prompt, reaches 73.2\% w2v ECA and 70.0\% emotion2vec ECA. This shows that LoRA-based task adaptation contributes to the gain, but remains insufficient for strong emotion control without explicit emotion-aware conditioning.

Despite the large emotion gains, the proposed method preserves speaker identity, quality, and intelligibility at levels comparable to the VEVO backbone. Apart from a slight drop in naturalness (UT-MOSv2 $3.060\rightarrow2.960$), speaker similarity, DMOS, and WER remain on par or improve, suggesting that explicit emotion control does not destabilize the underlying voice conversion.


\noindent\textbf{Subjective Evaluation}. For MOS, the proposed method shows comparable quality to VEVO ($4.018$, 95\% CI $[3.644, 4.392]$ vs.\ $3.878$, 95\% CI $[3.464, 4.292]$), with overlapping confidence intervals, suggesting that the overall perceptual quality is maintained. Figure~\ref{fig:listener} shows a larger gap in perceptual preference: the proposed method is preferred over VEVO in both emotion similarity ($75.2\%$ vs.\ $17.5\%$) and speaker similarity ($58.7\%$ vs.\ $16.8\%$). The Wilson 95\% CIs for the proposed preference rates are $[70.8, 79.2]\%$ for emotion similarity and $[53.9, 63.5]\%$ for speaker similarity. As a model-independent listening test, this subjective evaluation provides perceptual support for the objective trends and suggests that stronger emotion control does not come at the cost of perceived speaker similarity.

\subsection{Probing the Emotion and Style Pathways}
Beyond the AR-LoRA-only variant discussed above, the remaining ablation rows in the lower block of Table~\ref{tab:main_results} probe the two reference pathways of the Emotion-Aware Prefix encoder. Rather than removing modules, we feed the \emph{source}-emotion reference to a single pathway while keeping the architecture intact, which suppresses that pathway's access to target-emotion information. \textit{w/o Target Emotion Embedding} feeds the source reference to the emotion-fusion pathway ($\mathcal{E}$), and \textit{w/o Target Style Embedding} feeds it to the temporal-shuffle/style pathway ($s$). \textit{w/o Deep-Prefix} is instead an architectural variant that prepends $\mathbf{E}$ to the input rather than injecting it as layer-wise KV-cache.

The two probes affect different aspects of the output. Feeding the source reference to the emotion-fusion pathway reduces emotion accuracy to near the baseline level (ECA $91.2\%\rightarrow43.4\%$) while speaker similarity is largely retained (Spk-Cent SIM $0.500\rightarrow0.491$), suggesting that this pathway carries most of the target-emotion information while having limited impact on identity in this probe. Feeding the source reference to the style pathway instead leaves emotion largely intact (ECA $86.1\%$) but yields the largest drop in speaker similarity (Spk-Cent SIM $0.500\rightarrow0.464$), indicating that the style pathway also carries identity-relevant information, consistent with the speaker/style compression role of similar Perceiver-style encoders such as GenVC. Together, these observations suggest that the two pathways influence emotion and identity to different degrees, rather than encoding a single shared factor. Finally, \textit{w/o Deep-Prefix} causes only a small, consistent drop (ECA $91.2\%\rightarrow90.6\%$), indicating that layer-wise injection is a minor but reliable refinement over simple prepending.

\subsection{Where and How Is Emotion Controlled?}
To localize emotion control across the two stages, we prompt the target emotion to each stage in isolation: \textbf{Acoustic Only} prompts [\textcolor{blue}{source}, \textcolor{red}{target}], \textbf{Sequence Only} prompts [\textcolor{red}{target}, \textcolor{blue}{source}], and \textbf{Joint} prompts [\textcolor{red}{target}, \textcolor{red}{target}] to Stage~1 and Stage~2, respectively. Table~\ref{tab:hierarchical_sensitivity} and Figure~\ref{fig:confusion} reveal three main patterns.

\noindent\textbf{Sequence modulation is the primary driver of controllability}. For the proposed method, Sequence-Only control reaches $54.0\%$ ECA, substantially higher than Acoustic-Only control ($33.0\%$), indicating that Stage~1 carries much of the high-level prosodic intent needed for explicit emotion conversion. In contrast, VEVO shows the opposite pattern (Acoustic-Only $30.0\%$ vs.\ Sequence-Only $12.0\%$), suggesting that its original sequence modulation stage has limited access to target-emotion control.

\noindent\textbf{Acoustic prompting provides partial but insufficient control}. 
Acoustic-Only prompting can shift some categories such as \emph{Angry}, \emph{Happy}, and \emph{Sad} toward the target emotion, but the effect remains incomplete and uneven. In particular, \emph{Surprise} is poorly recovered, suggesting that acoustic cues alone are insufficient for emotions requiring stronger temporal and prosodic restructuring.

\noindent\textbf{Joint control aligns sequence intent with acoustic realization}. Joint control yields the highest ECA ($91.2\%$), far exceeding either stage alone. The confusion matrices show that all five emotions, including \emph{Surprise}, reach high accuracy under the proposed method, suggesting that Stage~1 specifies the target emotional trajectory while Stage~2 realizes it with target-speaker acoustic characteristics. In contrast, VEVO under Joint control still shows systematic confusion, consistent with a backbone that can exploit acoustic prompts but lacks explicit sequence-level emotion steering.

\subsection{The Role of Acoustic Decoupling in Identity Preservation}
To examine the structural role of acoustic decoupling, we apply the Emotion-Aware Prefix to GenVC (GenVC w/ EAP). As shown in Table~\ref{tab:genvc_comparison}, emotion improves substantially (ECA $32.2\%\rightarrow60.2\%$), but speaker identity collapses (EER $20.9\%\rightarrow44.5\%$) even under same-speaker prompting. This contrasts sharply with applying EAP to VEVO, which preserves identity (Table~\ref{tab:main_results}). 
One plausible explanation is structural: VEVO keeps a separate acoustic realization stage with a frozen FM Transformer, whereas GenVC feeds sequence-level representations directly into a fine-tuned vocoder. In this setting, strong sequence-level emotion steering appears to perturb waveform generation and speaker identity more severely. This comparison suggests that acoustic decoupling may play an important role in preserving speaker identity under aggressive emotion control.

\begin{table}[t]
\centering
\caption{Comparative Analysis: GenVC With and Without the Emotion-Aware Prefix.}
\label{tab:genvc_comparison}
\setlength{\tabcolsep}{8pt}  
\begin{tabular}{l|c|cc}
\toprule
\textbf{Model} & \textbf{EER} $\downarrow$ & \textbf{Emo SIM} $\uparrow$ & \textbf{ECA} $\uparrow$ \\ 
\midrule
GenVC & \textbf{20.87\%} & 0.321 & 32.2\% \\
\textbf{GenVC w/ EAP} & 44.51\% & \textbf{0.575} & \textbf{60.2\%} \\
\bottomrule
\end{tabular}
\end{table}



\section{Limitations and Future Work}
Our method assumes access to a reference of the \emph{target} speaker in the \emph{target} emotion to prompt the acoustic realization stage. This assumption holds in the standard reference-based EVC setting, but may not hold when only neutral or emotion-mismatched target-speaker references are available. In this stricter case, the system falls back to sequence-only control, where ECA drops from 91.2\% to 54.0\% (Table~\ref{tab:hierarchical_sensitivity}). This is consistent with our finding that sequence modulation is the primary driver of emotion control, while acoustic cues still provide a non-trivial contribution.

Developing models that can control emotion without target-emotion acoustic prompts while preserving speaker identity remains an important direction for future work. This may require better disentanglement of speaker identity, emotion, and acoustic realization, especially in settings involving unseen speakers or unseen target-emotion references.
\section{Conclusion}


This paper introduced the Emotion-Aware Prefix for explicit emotion control in voice conversion, improving Emotion Conversion Accuracy (ECA) from 47.5\% to 91.2\% while maintaining linguistic content, speech quality, and speaker identity. Our analyses show that joint control across sequence modulation and acoustic realization is important for reliable emotional expression. Comparative results further suggest that the proposed module transfers across backbones, while acoustic decoupling may help preserve speaker identity under strong emotion control.

\section{Generative AI Use Disclosure}

The authors used OpenAI's ChatGPT and Anthropic's Claude to polish and edit the manuscript. The authors also used Cursor's AI coding assistant to assist with small portions of code for experimental analysis and visualization. All research ideas, experimental designs, experiments, analyses, and conclusions were conceived, carried out, and validated by the authors. The authors reviewed and verified all AI-assisted outputs, including AI-assisted code, and take full responsibility for the content of this article.

\bibliographystyle{IEEEtran}
\bibliography{EVCbibtrimed}

\end{document}